# IET Radar, Sonar & Navigation

## Special Issue Call for Papers

**Be Seen. Be Cited.
Submit your work to a new IET special issue**

Connect with researchers and experts in your field and share knowledge.

Be part of the latest research trends, faster.

**Read more**

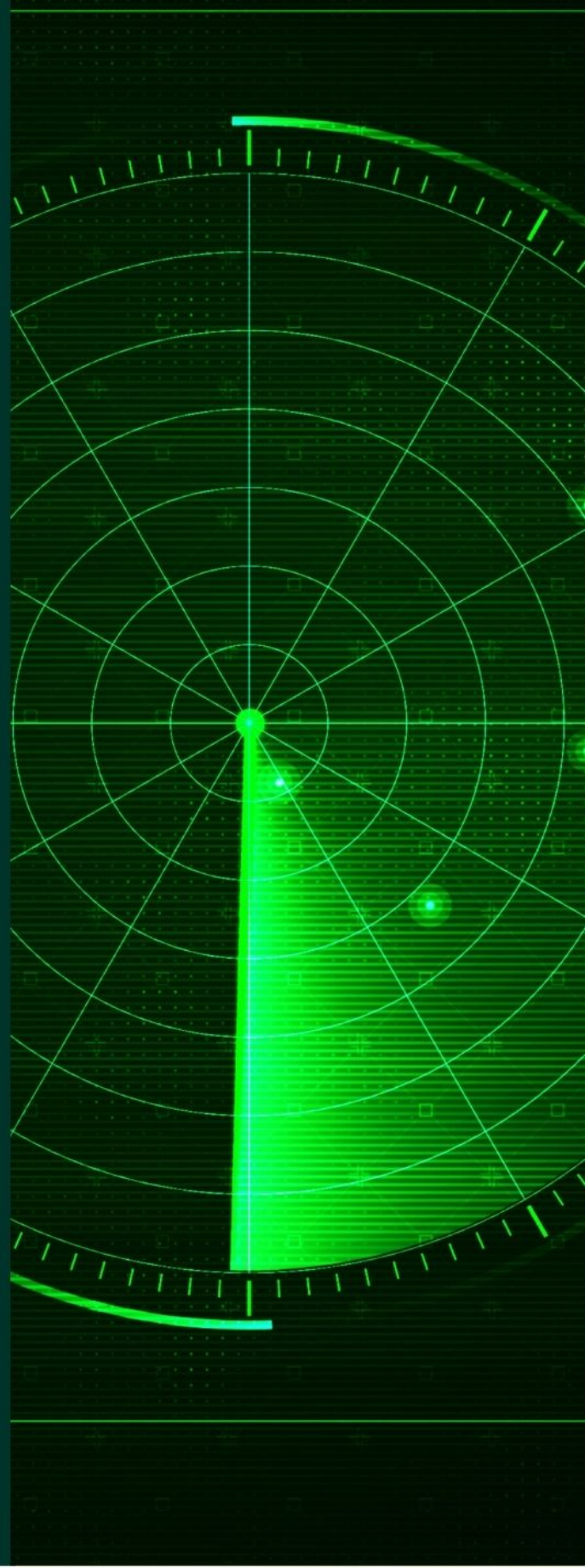
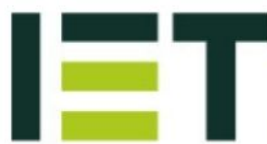

The Institution of Engineering and Technology

# One-shot learning-based driver's head movement identification using a millimetre-wave radar sensor

Hong Nhung Nguyen[1,2] | Seongwook Lee[3] | Tien-Tung Nguyen[4] | Yong-Hwa Kim[5]

[1]Department of Electronic Engineering, Myongji University, Yongin, Korea

[2]Faculty of Information Technology, Viet Tri University of Industry, Viet Tri, Vietnam

[3]Electronics and Information Engineering, College of Engineering, Korea Aerospace University, Goyang-si, Republic of Korea

[4]Faculty of Electronics Technology, Industrial University of Ho Chi Minh City, Ho Chi Minh City, Vietnam

[5]Department of Data Science, Korea National University of Transportation, Uiwang-si, Republic of Korea

**Correspondence**

Yong-Hwa Kim, Department of Data Science, Korea National University of Transportation, Uiwang-si, Gyeonggi-do, 16106, Republic of Korea.
Email: yongkim@ut.ac.kr

**Funding information**

National Research Foundation of Korea (NRF), Grant/Award Number: 2019R1A2C2086621; Korea National University of Transportation, Grant/Award Number: 2021

**Abstract**

Concentration of drivers on traffic is a vital safety issue; thus, monitoring a driver being on road becomes an essential requirement. The key purpose of supervision is to detect abnormal behaviours of the driver and promptly send warnings to him/her for avoiding incidents related to traffic accidents. In this paper, to meet the requirement, based on radar sensors applications, the authors first use a small-sized millimetre-wave radar installed at the steering wheel of the vehicle to collect signals from different head movements of the driver. The received signals consist of the reflection patterns that change in response to the head movements of the driver. Then, in order to distinguish these different movements, a classifier based on the measured signal of the radar sensor is designed. However, since the collected data set is not large, in this paper, the authors propose One-shot learning to classify four cases of driver's head movements. The experimental results indicate that the proposed method can classify the four types of cases according to the various head movements of the driver with a high accuracy reaching up to 100%. In addition, the classification performance of the proposed method is significantly better than that of the convolutional neural network (CNN) model.

**KEYWORDS**
convolutional neural nets, driver information systems, learning (artificial intelligence), pattern classification, radar signal processing

## 1 | INTRODUCTION

Nowadays, an explosive growth of smart solutions together with information technology is being witnessed in modern transportation systems. In particular, the successes of smart solutions depend critically on the role of advanced sensors. As a result, car manufacturers are developing sensors to deploy in vehicles for a variety of applications purposes, for example, safety, traffic management, and infotainment [1]. One of the urgent issues not only for individuals but also for car manufacturers to be solved is safe driving due to the increasing number of vehicles on roads. Safe driving is heavily dependent on the driver's behaviours to avoid traffic accidents. For instance, a timely warning sent to the driver who is considered sleepy will be able to prevent accidents. The abnormal signs related to a drowsy state of the driver such as eye closure, blink frequency, nodding frequency, face position, fixed gaze, heart rate, or erratic steering was investigated in Ref. [2]. Several methods such as vehicle-based measures [3]; behavioural measures [4] and physiological measures [5] have been performed to determine whether a driver is drowsy. In addition, the movement of the driver's head has been analysed to infer the driver's state of drowsiness [6, 7]. Human head movements were monitored using wearable sensors [6, 8, 9], cameras [10] and radars [11, 12]. The method of using a wearable sensor has the disadvantage of being cumbersome. The camera sensor is capable of recording, which is a privacy issue, and lighting conditions also have a negative impact on monitoring performance. Radars can monitor human head movements in light conditions and overcome the driver privacy issue [11, 12].

Recently, artificial intelligence in general and deep learning, in particular, have achieved success in a variety of applications such as intelligent chatbots, self-driving cars, virtual assistants, speech and image recognition [13]. Taking advantage of deep learning and in order to detect the drowsy state of the driver [11, 12], using the radar to collect data and apply the deep learning method to design a framework to monitor human head motion, the authors in Ref. [11] utilised a convolutional neural network (CNN) to classify four cases of the driver's head movements. In Ref. [12], a CNN was used to classify the eight head motions of the driver's head. However, the supervised learning-based approaches require a huge amount of data to be labelled.

To solve the problems of data scarcity, One-shot learning has been emerged as an effective tool applied to various research fields [14–20]. One-shot image classification based on a variational Bayesian framework was proposed with the premise that the previously learned classes could help predict future ones when there are very few examples available from a provided class [14]. In Ref [15], a path planning problem for robotic actuation was addressed via the One-shot learning method. In Ref [16], the One-shot learning method is used to classify handwritten characters based on the image dataset representing those handwritten characters. Interestingly, in Ref [17], the performances of One-shot learning-based prediction methods in drug discovery applications were evaluated. To address the issue of limited data, classifier-enabled few-shot learning for rolling bearing fault diagnosis was proposed in Ref [18]. One-shot learning aims to find information about object types from one or just a few training samples in Ref. [19]. In Ref. [20], the authors successfully used One-shot learning to classify partial discharges.

In our study, we focussed on detecting those alerts drivers when their head movements change; only a normal head position of the driver should be set for safe driving; the rest are abnormal positions. We propose One-shot learning to classify driver's head movements using the mmWave radar sensor. In our experiment, first, the radar sensor that was put in the centre of the steering wheel collected the reflected signals for four cases corresponding to the driver's head movements as the driver staring at the front, the driver shaking his/her head up and down, the driver shaking his/her head side to side, and the driver lowering her/his head. Then, we design a classifier to analyse the received radar signals and detect abnormal behaviour of the driver's head movement. The performance of our One-shot learning method is better than that of the CNN proposed in Ref. [11] because the proposed method is more optimised and uses more layers including batch normalisation as well as a dropout layer. The results of the proposed method have shown good efficiency compared to previous studies. The main contributions of this study can be summarised as follows:

- To the best of our knowledge, One-shot learning is demonstrated for the first time to classify driver's head movements based on the data collected from the small-sized 61 GHz frequency-modulated continuous-wave (FMCW) radar sensor.
- The proposed model uses the distance metric to map the data of radar-based driver's head movements and becomes an effective classifier for distinguishing the abnormality of the driver's head movements. This method employs pairs of the samples of the same or different classes during the training phase and classifies the test sample with a single training sample for each class. The proposed One-shot learning for the monitoring of driver obtains better classification performance than CNN and achieves a classification performance of almost 100%.

The remainder of the paper is structured as follows. In Section 2, we introduce basic radar signal processing and radar data acquisition methods. In addition, a method for converting radar signal data into a format for the One-shot learning is described. In Section 3, we present the structure of the One-shot learning model, which is based on the Siamese network and applied to radar data sensors. In Section 4, the classification results of the proposed method are presented. Finally, we conclude the paper in Section 5.

## 2 | MILLIMETRE WAVE FMCW RADAR

### 2.1 | Signal analysis for the FMCW radar system

In the FMCW radar system, a sequence of signals is transmitted whose frequency changes linearly with a function of time [21].

Figure 1 shows the time-frequency slope of the transmitted signal. The duration of a frame is set as 50 ms, which is divided into 12.5 ms for transmission duration and 37.5 ms for signal processing duration. In addition, the sampling frequency is 2 MHz and 256 samples are obtained from each transmitted signal. Table 1 presents the radar parameters and their values used in our measurement.

Each frame consists of a transmission and the corresponding signal processing duration.

The transmitted signal can be expressed as

$$f(t) = \cos\left(2\pi f_c t + \pi \frac{\Delta B}{\Delta T} t^2\right), \quad (1)$$

where $f_c$ is the carrier frequency. In addition, $\Delta B$ and $\Delta T$ are the operating bandwidth of the transmitted signal and the duration of transmission in one frame, respectively. The transmitted signal $f(t)$ is received back with a time delay and amplitude attenuation and it can be expressed as

$$r(t) = \alpha f(t - t_{td})$$
$$= \alpha \cos\left(2\pi f_c(t - t_{td}) + \pi \frac{\Delta B}{\Delta T}(t - t_{td})^2\right), \quad (2)$$

where $\alpha$ is the attenuation factor owing to the path loss, and $t_{td} = \frac{2R}{c}$ denotes the time delay. Additionally, $R$ is the distance to the target, and $c$ is the propagation velocity of the radar signal. Then, the transmitted and received signals pass through the frequency mixer, and high and low-frequency signals are generated. Finally, the output of the frequency mixer is passed through the low-pass filter to extract a baseband signal (i.e. a low-frequency signal). The baseband signal can be expressed as

$$m(t) \cong \frac{\alpha}{2} \cos 2\pi \left(\frac{2\Delta BR}{\Delta Tc} t + \frac{2f_c R}{c}\right). \quad (3)$$

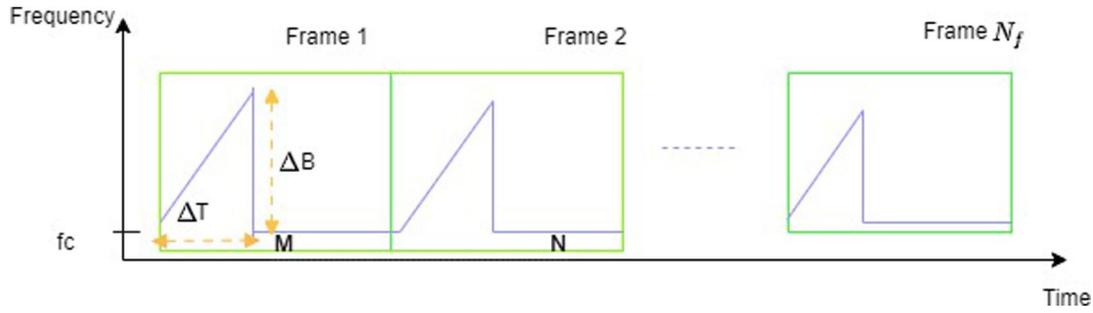

**FIGURE 1** Time-frequency slope of the transmitted frequency-modulated continuous-wave (FMCW) radar signal

**TABLE 1** Parameters used in the radar system

| Parameters | Value |
| --- | --- |
| The duration of transmission | 12.5 ms |
| Wait time | 37.5 ms |
| Analog-to-digital converter sampling frequency | 2 MHz |
| Maximum detectable range | 2 m |
| Range resolution | 2.5 cm |
| Velocity resolution | 0.7 km/h |
| Centre frequency | 61 GHz |

The baseband signal is also referred to as a beat signal because the frequency difference between $f(t)$ and $r(t)$ is contained in this signal. After a sampling process, the beat signal $m(t)$ becomes

$$m[n] \cong \frac{\alpha}{2} \cos\left(2\pi \frac{2\Delta BR}{\Delta Tc} nT_s + 2\pi \frac{2f_c R}{c} + \phi\right) \quad (4)$$
$$(n = 0, 1, \ldots, N_s - 1),$$

where $T_s$, $N_s$, and $\phi$ are the sampling period, the number of samples, and the phase offset because of the sampling, respectively. Finally, the fast Fourier transform is used to extract the beat frequency of the time-sampled signal in the FMCW radar system. The resulting frequency-domain beat signal is denoted as

$$M[k] = \sum_{n=0}^{N_s-1} m[n] \exp\left(\frac{-j2\pi kn}{N_s}\right) \quad (5)$$
$$(k = 0, 1, \ldots, N_s - 1).$$

In this signal, a peak value appears at the index $k$ corresponding to the target. From that peak value, we can estimate the distance to the target as

$$\hat{R} = \frac{c\Delta T}{2\Delta B} f_b. \quad (6)$$

## 2.2 | Measurement environment

Figure 2a shows the experiment environment for measuring the radar signals. To obtain data, we used a millimetre wave FMCW radar sensor made by bitsening Inc. as shown in Figure 2b, which operates at a carrier frequency of 61 GHz and a bandwidth of 6 GHz. Here, we can see that the radar is set up at the centre of the steering wheel and facing the driver's head. This position provides the radar to catch an expansive illustration of the driver's range of movement. In addition, the distance between the driver's head and the steering wheel is about 40 cm. Hence, we crop the radar image to utilise only the signals reflected from 20–60 cm.

First of all, the frequency band of the radar system we used is a 60 GHz band and does not overlap with the frequency band currently used by 4G/5G communication systems. We also applied a mean subtraction method [22] to the raw radar signal to remove clutter reflected from structures fixed inside the vehicle. Therefore, the adverse effect of multi-path reflection was mitigated.

The radar is downsized using the high frequency. The range resolution is inversely proportional to the bandwidth

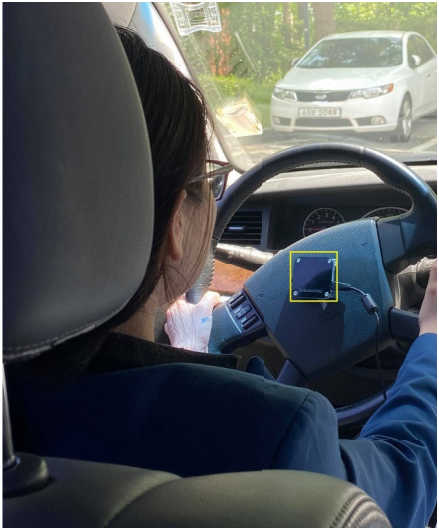

**(a)** Experimental environment in a car

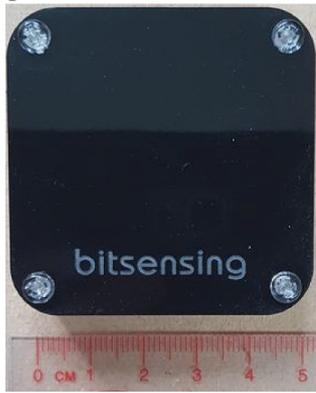

**(b)** Millimeter-wave radar

**FIGURE 2** Measurement environment

[23]; therefore, a wide bandwidth allows high detection performance. The radar was installed on the middle of the steering wheel to face the driver's head. As shown in Figure 3, we acquired the radar signals for four different types of head movements: (Case 1) the driver is staring at the front, (Case 2) the driver is shaking his head up and down, (Case 3) the driver is shaking his head from side to side, and (Case 4) the driver is lowering his head. On all types of head movements of three drivers, experiments were conducted and the radar data were collected based on four cases of driving scenarios.

## 3 | PROPOSED METHOD FOR MONITORING DRIVER'S HEAD MOVEMENT

In this section, we describe input-signal design and the proposed method to classify radar-based driver head movements based on the One-shot learning approach.

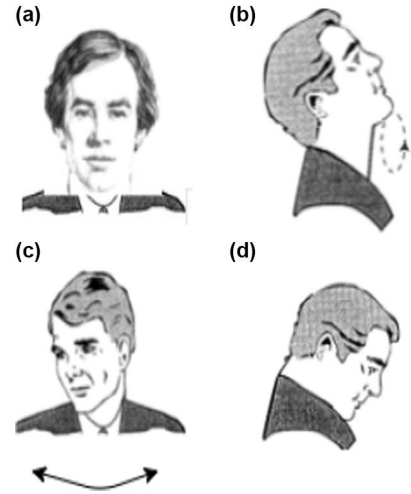

**FIGURE 3** Four types of driver's head movements: (a) Case 1: when staring at the front; (b) Case 2: when shaking head up and down; (c) Case 3: when shaking head side to side; and (d) Case 4: when lowering the head

### 3.1 | Designing input from the FMCW radar signal

Figure 4 illustrates the frequency spectrum of the beat signal in a single frame when the distance between the driver's head and the radar is 40 cm. The frequency of the real-valued signal is symmetric with respect to its direct current component; therefore, we use the first half of the samples.

$$\mathbf{M} = \left[M[0], M[1], \cdots, M\left[\frac{N_s}{2} - 1\right]\right]^T. \quad (7)$$

Furthermore, the sample with index $k$ is converted to distance by the below equation,

$$R = \frac{c\Delta T}{2\Delta B} \times \frac{k}{N_s} f_s. \quad (8)$$

We use the matrix form signal

$$\mathbf{X} = \begin{bmatrix} \mathbf{M}^{(1)} \\ \mathbf{M}^{(2)} \\ \vdots \\ \mathbf{M}^{(N_f)} \end{bmatrix}^T, \quad (9)$$

where $\mathbf{M}^{(k)}$ is the frequency spectrum in Equation (7) corresponding to the $k$th frame, and $N_f$ is the number of frames to be observed. In our measurement, we use 40 points that correspond to a length of 1 m and also use 30 frames corresponding to the 1.5 s observation time. Thus, we have that $\mathbf{X} \in \mathbf{R}^{N_f \times N_s'}$ with $N_f = 30$ and $N_s' = 40$.

A different signal pattern can be obtained for each driver's head case. Figure 5 shows the input data in Equation (9) for

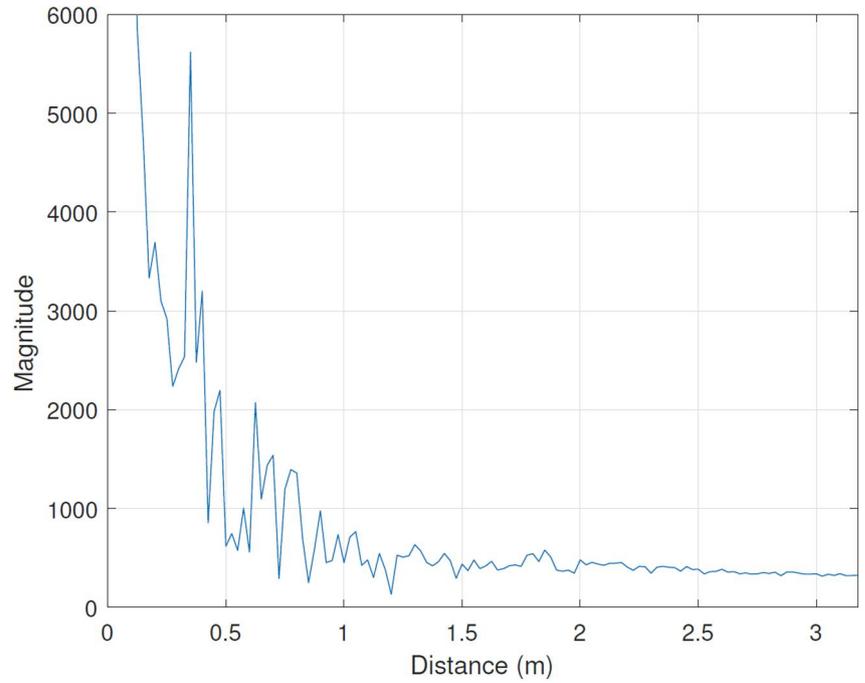

**FIGURE 4** Frequency spectrum of the beat signal in a single frame

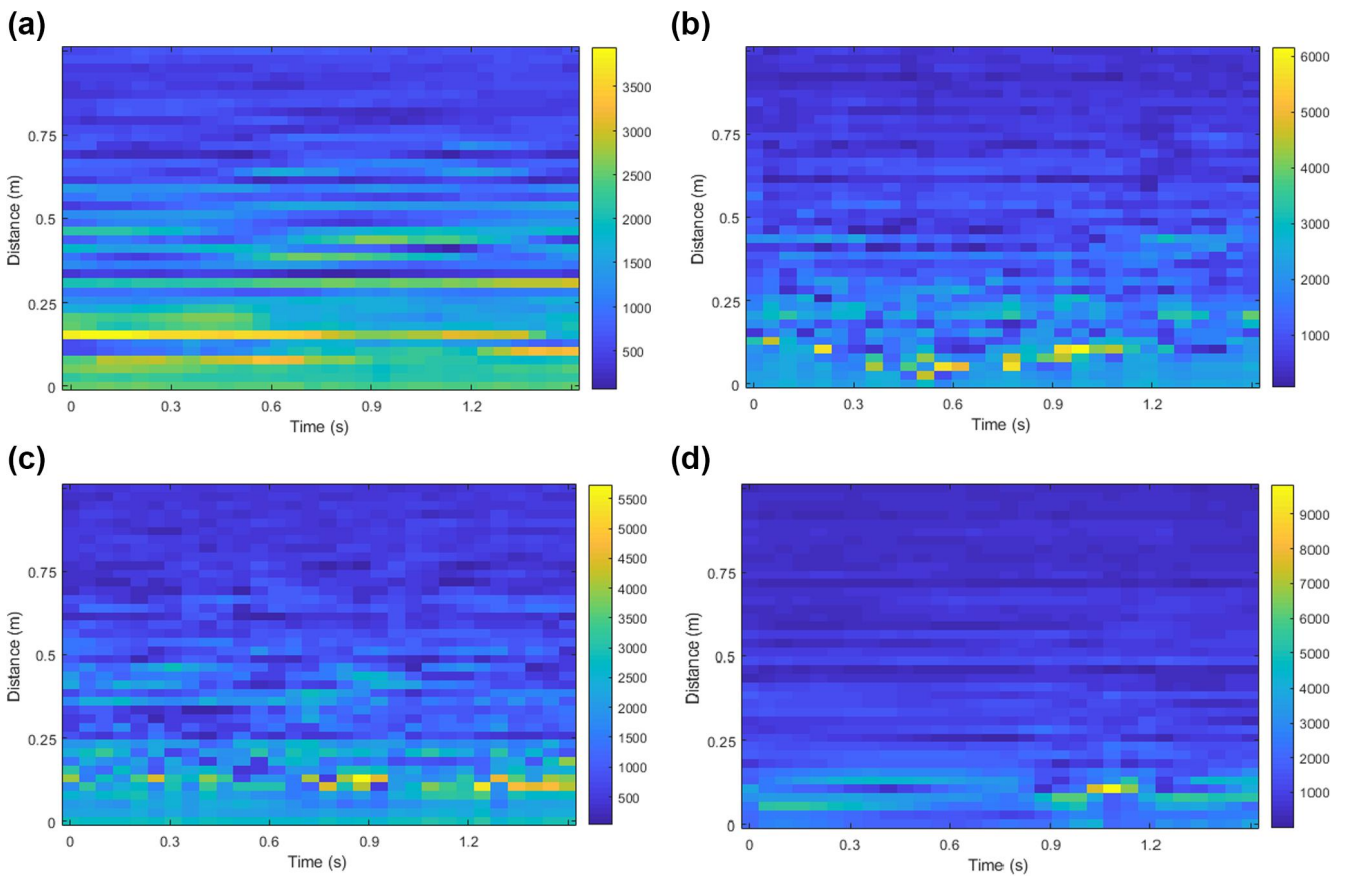

**FIGURE 5** Illustration of four types of driver's head movements based on the radar signal: (a) Case 1: staring at the front; (b) Case 2: shaking head up and down; (c) Case 3: shaking head side to side; and (d) Case 4: lowering the head

the proposed model from the radar signal. It can be easily seen that there are two pairs, that is, Figure 5a,d presenting Case 1 and Case 4, respectively, and Figure 5b,c presenting Case 2 and Case 3 having relatively similar colour lines. Note that when the driver's head moves less or is stationary, the measured signal from the sensor changes less as shown in

Figure 5a,d (Case 1 and Case 4). However, the position of the driver's head also affects the measured signal, which makes a difference of colour lines between Case 1, that is, the driver is staring at the front, and Case 4, that is, the driver is lowering the head. In cases wherein the driver's head moves more as in Case 2, that is, the head of the driver is shaking up and down, and Case 3, that is, the driver's head is shaking side to side, the colour lines become more unsettled as in Figure 5b,c. Furthermore, the colour lines of Case 2 and Case 3 are slightly different due to the change in the position of the driver's head. We collected radar signals for four different head movements. The first case is the driver's standard and the others correspond to abnormal behaviour. Moreover, the periodic frequency is observed for the motions in Cases 2 and 3. The data obtained in Cases 2 and 3 is moving data (the driver is shaking his head up and down and the driver is shaking his head from side to side) and the data in Cases 1 and 4 is static data (the driver is staring ahead and the driver is lowering his head). Therefore, a more similar colour line is observed for the stationary motion in Cases 1 and 4.

## 3.2 | One-shot learning method

The proposed method for classifying driver's head movements based on the data of the radar sensor is performed by One-shot learning. Figure 6 illustrates the architecture of the proposed method based on a One-shot learning model. As shown in the figure, the method consists of two phases, that is, training and test phases. In addition, the dataset is divided into three parts, namely, training set $\mathcal{T}$, test set $\mathcal{N}$, and the support set $S_S$. In the training phase, the data input consists of many sample pairs, each of which is a sample pair with the same or different classes, and the output of the model is the distance to evaluate whether the sample pair is the same for different classes.

In the test phase, a scenario is denoted with a support test $S_S$ containing $K$ labels and $N$ samples. We propose that each class employed as a support sample is provided by a single-data using a radar sensor ($N = 1$). that is, One-shot learning classification for driver's head movements using the radar sensor.

For more details, Figure 7 shows the One-shot learning model using a Siamese neural network for classification of driver's head movement based on a 61 GHz radar sensor.

Our model with the Siamese neural network contains twin CNN networks that accepts distinct inputs, and it is connected by an energy function that figures out some metrics between the highest level feature representations on each side and the outputs are combined to provide some prediction [24, 25].

As can be seen in Figure 7, working in parallel, each of these inner networks receives an input vector based on which an output vector is generated. These output vectors can then be compared to see how similar they are. Intuitively instead of trying to classify inputs to one of the four head's positions, a Siamese network learns to differentiate between inputs, learning their similarity and denoting the chances that the two input images belong to the same head's movement. As a result, when training the Siamese network, we need to have the same or different pairs, which are randomly sampled in the training model. The same pair means two images belong to the same class (e.g. two samples of the same instance of the head's driver). In contrast, different pairs indicate two images are of different classes. After the training has been performed, the Siamese network is capable of telling us if two new images are similar enough to be the same class or not.

We denote the input of the twin sub-networks CNN as a pair with the same or different classes $(\mathbf{X}_1, \mathbf{X}_2)$ where $(\mathbf{X}_1, \mathbf{X}_2)$, which are two signals presented in Equation (9). The distance metric between their output on the twin sub-networks is calculated as

$$\mathbf{d}_f^2(\mathbf{X}_1, \mathbf{X}_2) = \|f(\mathbf{X}_1) - f(\mathbf{X}_2)\| \qquad (10)$$

where $f(\mathbf{X}_1)$ and $f(\mathbf{X}_2)$ represent the feature vectors extracted by the twin sub-network CNN.

Basically, the two CNN networks have the same parameters and weights. Each CNN consists of convolutional layers, dropout, flatten, and fully connected layers as shown in Figure 8. The filter number of convolution layers is set in multiples of 16 to optimise performance, and the output feature map of each convolution in the first three layers is applied to the Rectified Linear Unit (ReLU) activation function, which is currently one of the most common activation in the networks [26]. The dropout layers, which help prevent overfitting of the model are added for regularisation [27] and batch normalisation is used to speed up learning by normalising the input of the convolution layer [28]. Batch normalisation helps the learning process and dramatically decreases the number of training epochs needed to train deep networks [29]. Before or after the activation function in the previous layer, batch normalisation may be applied to the inputs to the layer. In our experiment, we examined appending batch normalisation before and after ReLu activation. The results show that the performance of the model are better when the batch normalisation layer is added after the ReLu activation.

It has been mentioned that CNN can learn features on its own. The most remarkable characteristic of CNN is that the weights of the filters are retained automatically instead of manually, while the network gains high accuracy. Using the convolutional layer, each hidden unit is associated with one block of the input image and extracts features from that block, leading to the construction of the feature map. The feature maps capture the result of applying the filters to an input image at each layer; the feature map is the output of that layer [30]. The feature map is converted from a 2-dimensional matrix into a single vector by using a flatten layer. Finally, the fully connected layer is applied in the network.

The output is the feasibility distance of the feature vector outputs from sub-network CNN twins as expressed in the following expression:

$$P(\mathbf{X}_1, \mathbf{X}_2) = \sigma(FC(\mathbf{d}_f^2(\mathbf{X}_1, \mathbf{X}_2))), \qquad (11)$$

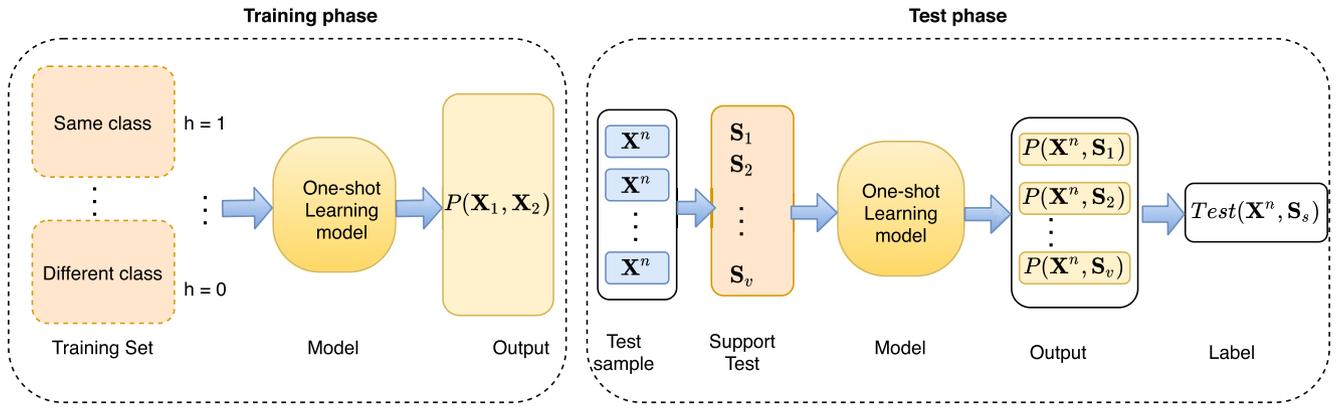

**FIGURE 6** Flowchart of training and testing in One-shot learning

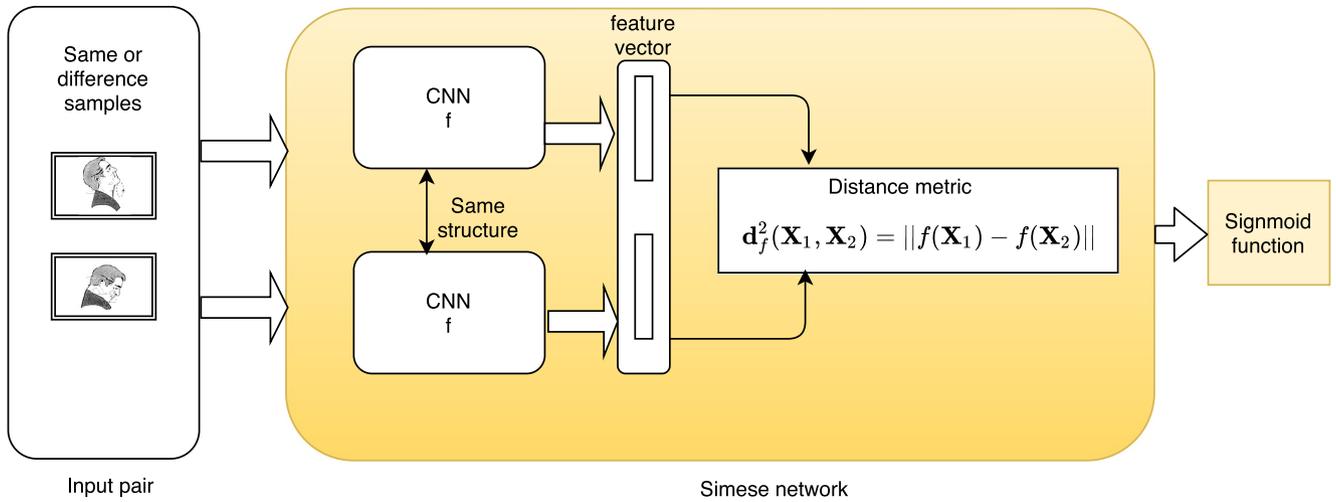

**FIGURE 7** One-shot learning model using the Siamese network

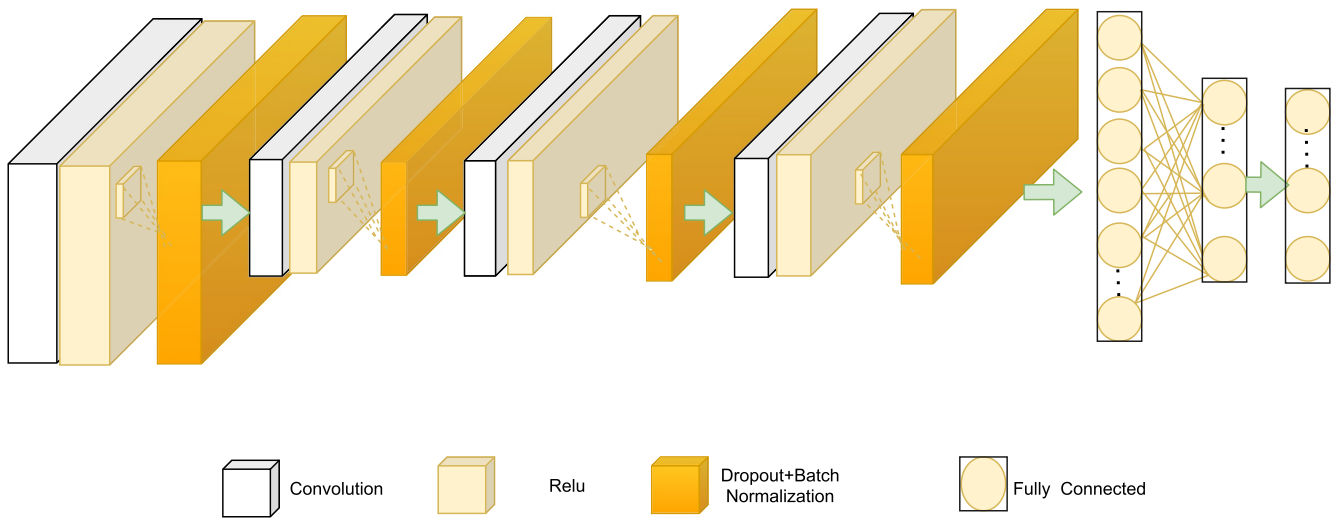

**FIGURE 8** Structure of the convolutional neural network (CNN) used in the proposed method

where $\sigma(\cdot)$ is the sigmoid function and $FC$ is a fully connected layer.

## 3.3 | Network optimization

In the training phase as shown in Figure 6, the output is compressed into [0, 1] with the sigmoid function to translate it into probability. The output label is denoted as $h$, the target $h = 1$ is set when the radar sensor data $\mathbf{X}_1$ and $\mathbf{X}_2$ are in the same class, whereas $h = 0$ is set for a difference class.

The cross entropy between prediction and the target is used as a loss function, which is calculated as

$$Loss(\mathbf{X}_1, \mathbf{X}_2, h) = h \log(P(\mathbf{X}_1, \mathbf{X}_2)) \\ + (1-h) \log(1 - P(\mathbf{X}_1, \mathbf{X}_2)). \quad (12)$$

Various optimization algorithms exist, such as AdaGrad, AdaDelta, Nesterov and Adam Optimiser to minimise loss of the function [31–33]. We selected the Adam optimiser, which is a generalisation of the Adagrad algorithm by computing and updating statistics such as the first and second moments of the historical slope at each iteration. As shown in Figure 6, in the test phase, in order to evaluate the proposed networks, the dataset was expressed as difference classes that have not been exposed in the training phase. The trained model has no overfitting problems, which is one of advantages of testing on a new dataset. It is used as test samples $\mathbf{X}^n \in \mathcal{N}$ for classification, and the support test $S_S$, which includes $K$ samples $S_S = \{\mathbf{S}_1, ..., \mathbf{S}_K\}$ where $v = 1, ..., K$ and $\mathcal{N}$ share the same label set with the support set $\mathbf{S}_S$. The test samples are classified into a class as

$$Test(\mathbf{X}^n, \mathcal{S}_S) = \underset{v}{\operatorname{argmax}} P(\mathbf{X}^t, \mathbf{S}_v). \quad (13)$$

The performance of the proposed One-shot learning method is calculated as

$$\text{Accuracy} = \frac{\text{Count of } Test(\mathbf{X}^n, S_S) \text{ is correctly classified}}{|\mathcal{N}|}, \quad (14)$$

where $|\cdot|$ is the number of elements in a set.

## 4 | EXPERIMENTS AND RESULTS

In this section, we present performance of the proposed One-shot learning for the radar-based dataset to classify driver's head movements by using the radar signals we collected. We used TensorFlow and Keras [34] frameworks to build and develop our model. The model was trained and tested on an NVIDIA Titan X GPU with 3584 cores, each running at 1.4 GHz.

As mentioned in Section 3, in order to collect the radar signals, we used a millimetre-wave FMCW radar sensor with a distance between the driver's head and the radar being 40 cm. In this study, three people involving two men and one woman, aged between 24 and 42 years old participated in the experiments.

The case number in the experiments is shown in Table 2 together with the amount of data. The dataset was composed by a total of 5481 samples that contain the four considered cases of driver's head movements: (a) Case 1 with 1395 matrices for the driver staring at the front; (b) Case 2 with 1346 matrices for the driver shaking head up and down; (c) Case 3 with 1378 matrices for the driver shaking head side to side; and (d) Case 4 with 1362 matrices representing the driver lowering the head as shown in Figure 3. For convenience, numbers from 0 to 3 were assigned to the labels corresponding to each of the case driver's head movements.

Then, we divided the data into training and testing as follows: 72% for training and 8% for validation and 20% for testing. In order to achieve the other optimised hyperparameters, such as the number of epochs, batch size, and learning rate, we conducted extensive experiments with different parameters to adjust our model. The CNN model structure of the proposed architecture is detailed in Table 3, which contains name layers, output shapes, activation functions, kernel number and padding. The optimization step is performed in small batches of 64 samples and the learning rate was set to 0.006. Furthermore, after choosing the final parameters for the proposed model, to deal with the randomness of the algorithm, we performed the model up to 10 times with the same last parameters selected. Table 4 shows the performance classification and total parameter comparisons for the CNN model [11] and the proposed CNN and One-shot learning schemes, where the proposed CNN method has the same architecture as the proposed One-shot model, which was designed with 15 layers. Comparing both results, it can be concluded that the accuracy of One-shot learning is higher than that of the CNN. In particular, One-shot learning can reach an overall accuracy of 100% and a 1.94% higher accuracy than that of CNN, which has the same weights and parameters of CNN in the Siamese network used in the proposed method. This is because the proposed One-shot learning method studies distance data using two CNNs to map radar data into an appropriate embedded space and reduce variation in the class to avoid false identification. The authors in Ref. [11] utilised a CNN to classify four cases of the driver's head movements with performance of about 84.5%, and the accuracy of our proposed CNN is higher than approximately 13.5% compared to that of the CNN in Ref. [11].

This is because the structure of the proposed method has been added a number of layers and the structure was changed accordingly to improve its accuracy to more than that of the CNN in the previous paper [11]. As can be seen from Figure 8, in the structure of proposed method, the convolution layer is added with the number of filters, which is a multiple of 16 with kernel sizes of 3 × 3 and 5 × 5, 2 × 2. In order to bring down input size of the sample for the next convolution layers, we use the convolution layer with stride $S = 2$, which helps to reduce complexity and parameters in the network. In addition, each convolution layer is followed by batch normalisation [29] and a dropout layer with a learning rate drop factor of 0.5 has been applied to improve the performance of the neural network and help prevent overfitting [27, 35]. Due to the increase in the number of parameters, the classification accuracy of the proposed CNN rises significantly to reach 98.06%, while that of CNN [11] is 84.5%. In addition, comparing both the proposed techniques using the number of trainable parameters, it is clear that despite having an equal number of layers, the proposed CNN requires more learnable parameters than the proposed One-shot method. This is because while the SoftMax function of the CNN model has the role of predicting the probability distribution for the four motions of the head of the output layer, the One-shot method shares its parameters and outputs only 0 or 1 to classify the samples. The results demonstrated that our proposed method is effective in classifying driver's head movements using a radar-based sensor.

**TABLE 2** Experimental dataset

| Type of cases | Case 1(0) | Case 2(1) | Case 3(2) | Case 4(3) |
| --- | --- | --- | --- | --- |
| Number of samples | 1395 | 1346 | 1378 | 1362 |

**TABLE 3** Structure of the convolutional neural network (CNN) in the proposed method

| No. | Layer type | Kernel size/stride | Kernel number | Output size | Padding |
|---|---|---|---|---|---|
| 1 | Conv + ReLU | 2 × 2/1 | 16 | 40 × 30 × 16 | Same |
| 2 | Batch normalisation | - | - | 40 × 30 × 16 | - |
| 3 | Drop out | 0.5 | - | 40 × 30 × 16 | - |
| 4 | Conv + ReLU | 5 × 5/2 | 32 | 20 × 15 × 32 | Same |
| 5 | Batch normalisation | - | - | 20 × 15 × 32 | - |
| 6 | Drop out | 0.5 | - | 20 × 15 × 32 | - |
| 7 | Conv + ReLU | 5 × 5/1 | 64 | 20 × 15 × 64 | Same |
| 8 | Batch normalisation | - | - | 20 × 15 × 64 | - |
| 9 | Drop out | 0.5 | - | 20 × 15 × 64 | - |
| 10 | Conv + ReLU | 3 × 3/1 | 128 | 20 × 15 × 128 | Same |
| 11 | Batch normalisation | - | - | 20 × 15 × 128 | - |
| 12 | Drop out | 0.5 | - | 20 × 15 × 128 | - |
| 13 | Flatten | - | - | 38,400 | - |
| 14 | Fully connected | - | - | 64 | - |
| 15 | Fully connected | - | - | 32 | - |

**TABLE 4** Classification accuracy and total of parameters comparison

| Schemes | Accuracy | Total number of parameters |
|---|---|---|
| CNN [11] | 84.5% | 1396 |
| CNN in the proposed method | 98.06% | 2,598,612 |
| One-shot learning | 100% | 2,598,161 |

Figure 9a,b illustrate the confusion matrix results from the test set. In particular, Figure 9a shows that all of the test samples were correctly predicted. These results confirm that the performance of the proposed model affects the monitoring of the driver's head. Comparing the results between the CNN model and the proposed model, it is clear that the One-shot learning approach achieves higher accuracy than the CNN model in terms of performance classification. The One-shot learning-based classifier can predict driver's head movements with a surprising degree of accuracy reaching to 100%. Specifically, the classification accuracies are 99%, 95%, 99%, and 99% for Case 1, Case 2, Case 3, and Case 4, respectively. The average classification accuracy of CNN is 98%, whereas the One-shot learning model achieves a classification performance of almost 100%.

In order to evaluate the effect of the sample size on the performance of the proposed model, we conducted experiments with different sizes of training sets. Table 5 shows the results of classification accuracy of the proposed schemes with different sizes of the training dataset. To have a fair comparison between the proposed One-shot network architecture and the CNN, we also present the result of training both the models with the different volume of the dataset, that is, 10%, 20%, 30%, and 50% of samples of the entire data set, where all training hyper parameters and loss functions were kept the same as described before in this section for respective neural network models. As shown in Table 5, it is observed that although the CNN and the One-shot network have a competitive performance when trained on the large identical data, at 97.2% and 98.7% each with 50% samples of the whole data set used, the performance of the One-shot method is better than that of CNN at all scales of the training dataset. The results in this table also suggest that the performance gap between CNN and the One-shot method increases as there is a decrease in the size of the dataset used for its training. It can be seen that with 20% of the training samples using the driver's head motion classification accuracy rate of the CNN and the proposed model are 92.5% and 95.3%, respectively, while when only using 10% of the data set's samples, these models provide classification accuracies of 83.3% and 88.3%. Therefore, the proposed one-shot learning method has the potential to be implemented easily and quickly in real-world scenarios in the case of limited training data. With ever-growing safety demands and the increasing requirements for the driver's automation concentration in traffic, this can form a suitable application for situations where data annotation is difficult, or data availability is limited.

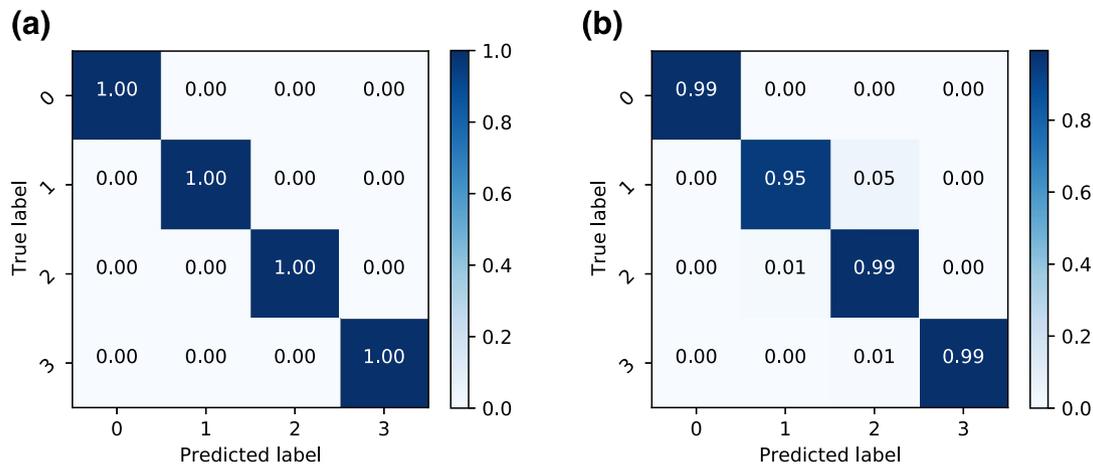

**FIGURE 9** Confusion matrix: (a) One-shot learning and (b) Convolutional neural network (CNN)

**TABLE 5** Performance evaluation with different scales of the dataset

| Rate of training dataset | Accuracy of the proposed CNN | Accuracy of the proposed One-shot |
| --- | --- | --- |
| 10% (548 samples) | 83.3% | 88.3% |
| 20% (1096 samples) | 92.5% | 95.3% |
| 30% (1644 samples) | 95.0% | 96.8% |
| 40% (2740 samples) | 97.2% | 98.7% |

In order to evaluate and better understand the effect of One-shot learning in the classification of radar-based data related to driver's head movements, we used the t-distributed Stochastic Neighbor Embedding (t-SNE) method, which is a tool for visualising high-dimensional data [36]. In principle, the t-SNE embeds high-dimensional vectors to 2-D spaces while retaining the pairwise similarity [36]. The t-SNE algorithm is only interested in the distance between the points; the algorithm locates the points on a plane. This paper uses a t-SNE method to visualise the data before and after training by the CNN model and proposed One-Shot learning method. Here, t-SNE has helped reduce the data dimension from multi-dimensional to only 2-dimensional space with change and visualise similar samples transformed into neighbouring points. Note that all samples of data, each of which has been featured in a 30 by 40 matrix form are as shown in Equation (9). Using the t-SNE algorithm, input data will be transformed into new expressions in the form of points and illustrated in Figure 10a. The points close to each other on the low-dimensional surface represent states that are similar in the high-dimensional space [36]. As shown in Figure 10a, it is noticed that the radar-based input signals for the four cases of driver's head movements are overlapped and remarkably close to each other. In the input data, Case 1 and Case 2 merged amalgamate with Case 3 and Case 4. Therefore, it is so difficult to classify all cases based on the radar-based input signals. Figure 10b,c show feature vector visualization via t-SNE for One-shot learning and for the CNN model in the last hidden fully-connected layer visualization, which is trained with 4384 samples, respectively. As shown in Figure 10b, which is drawn by the last hidden layer of the One-shot learning model, the radar-based signals have been stratified into distinct categories; Case 1 and Case 2 were completely separated, with no samples in Case 3 and Case 4 connecting them. Figure 10 visualises the feature vectors of the last hidden fully connected layer in the CNN model. As shown in Figure 10c, some instances in Case 2 became mixed with Case 1 and Case 2 with Case 4, respectively. It means that the data in the One-shot learning-based classifier are classified better than in the proposed CNN model.

To verify the proposed model results on more people, the data with three people studied. We measured the data by two more people in addition to the two already measured. After performing additional data measurements, the total dataset consisted of 16,681 samples which is shown in Table 6.

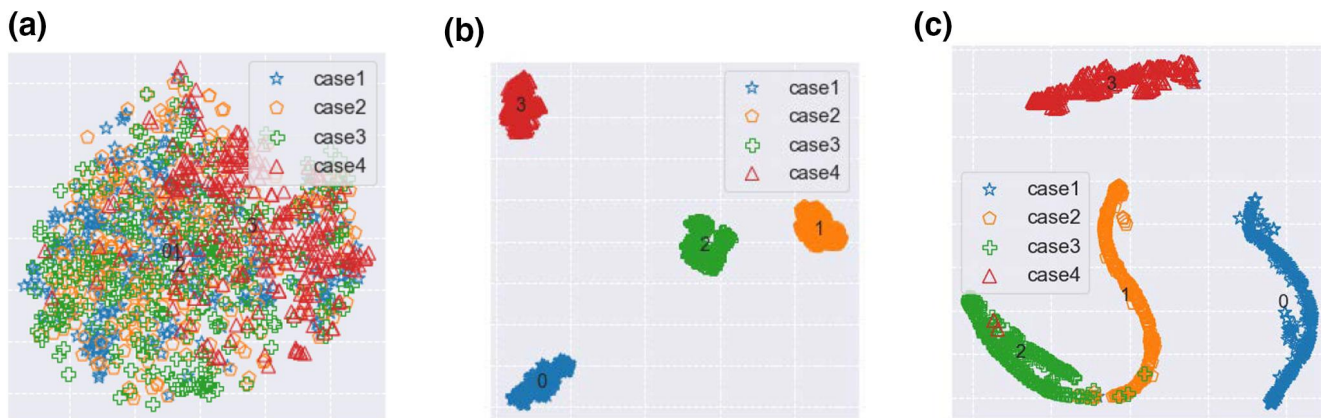

**FIGURE 10** Visualization of data (a) with input data, (b) with feature vector of the last hidden layer in One-shot learning, and (c) Convolutional neural network (CNN) model

**TABLE 6** Additional experimental dataset

| Type of cases | Case 1(0) | Case 2(1) | Case 3(2) | Case 4(3) |
| --- | --- | --- | --- | --- |
| Number of samples | 4195 | 4146 | 4178 | 4162 |

**TABLE 7** Performance evaluation with the additional dataset

| Data type | Proposed CNN model (%) | Proposed One-shot learning model (%) |
| --- | --- | --- |
| Case 1 | 100 | 100 |
| Case 2 | 99 | 100 |
| Case 3 | 100 | 100 |
| Case 4 | 100 | 100 |

We split the data into training and testing considering 80% for training and 20% for testing. After performing the model on the training dataset, we achieved high results for both the proposed CNN and the One-shot learning model. The test results obtained from the testing data are shown in Table 7.

The above results have proved the effectiveness of the proposed model. Along with the increased amount of data, the results of the CNN model are also increased compared to the limited amount of data measured from three people. However, the proposed one-shot model still gives a better result than the proposed CNN model. In the future, to apply more widely, we will practice testing on a more considerable amount of data with more people so that the model can be used more effectively in practice.

## 5 | CONCLUSION

In this paper, we proposed a One-shot learning approach based on Siamese neural networks to classify head movements of drivers for four cases using 61 GHz FMCW radar sensor signals. First, the radar signals were collected by performing various head movements and storing the frequency spectrum of the beat signal converted into an image. Subsequently, the proposed method extracts the radar signal features from two identical CNNs and measures the similarity of driver's head movements based on the distance metric. Our results indicated that the proposed method for radar-based driver's head movements' classification is expected to help prevent car accidents by detecting abnormal behaviours of drivers. In the future, we hope to conduct experiments with cars in motion states for practical applications and conduct further verifications of the proposed method using more datasets from many people.

### ACKNOWLEDGEMENT
This work was supported in part by the Korea National University of Transportation in 2021 and in part by the National Research Foundation of Korea (NRF) funded by the Korea government (MSIT) under Grant 2019R1A2C2086621. We thank Jaehoon Jung and Sohee Lim for sharing the radar sensor data they acquired.

### CONFLICT OF INTEREST
None.

### DATA AVAILABILITY STATEMENT
Data that support the findings of this study are available from the corresponding author upon reasonable request.

### ORCID
*Hong Nhung Nguyen*  https://orcid.org/0000-0002-2072-3973
*Seongwook Lee*  https://orcid.org/0000-0001-9115-4897
*Tien-Tung Nguyen*  https://orcid.org/0000-0001-7380-7591
*Yong-Hwa Kim*  https://orcid.org/0000-0003-2183-5085